\documentclass[runningheads]{llncs}

\usepackage[T1]{fontenc}

\usepackage{graphicx}
\usepackage[colorlinks=false]{hyperref}
\usepackage{booktabs}
\usepackage{multirow}
\usepackage{float}
\usepackage{array}
\usepackage{amsmath}

\begin{document}
\title{Enhancing JavaScript Malware Detection through Weighted Behavioral DFAs}

\author{Pedro Pereira\inst{1}\orcidID{0009-0008-7641-1566} \and
José Gonçalves\inst{2}\orcidID{0009-0004-1038-8384} \and
João Vitorino\inst{3}\orcidID{0000-0002-4968-3653} \and
Eva Maia \inst{4}\orcidID{0000-0002-8075-531X} \and
Isabel Praça \inst{5}\orcidID{0000-0002-2519-9859}}

\authorrunning{P. Pereira et al.}

\institute{GECAD, ISEP, Polytechnic of Porto, rua Dr. António Bernardino de Almeida, 4249-015 Porto, Portugal \\
\email{\{peesp,jpsgs,jpmvo,egm,icp\}@isep.ipp.pt}}
\maketitle              % typeset the header of the contribution
\begin{abstract}
This work addresses JavaScript malware detection to enhance client-side web application security with a behavior-based system. The ability to detect malicious JavaScript execution sequences is a critical problem in modern web security as attack techniques become more sophisticated. This study introduces a new system for detecting JavaScript malware using a Deterministic Finite Automaton (DFA) along with a weighted-behavior system, which we call behavior DFA. This system captures malicious patterns and provides a dynamic mechanism to classify new sequences that exhibit partial similarity to known attacks, differentiating them between benign, partially malicious, and fully malicious behaviors. Experimental evaluation on a dataset of 1,058 sequences captured in a real-world environment demonstrates the capability of the system to detect and classify threats effectively, with the behavior DFA successfully identifying exact matches and partial similarities to known malicious behaviors. The results highlight the adaptability of the system in detecting emerging threats while maintaining transparency in decision making.

\end{abstract}

\section{Introduction}
Nowadays, ensuring client-side security is a critical challenge \cite{client}, as attackers increasingly exploit vulnerabilities to manipulate, inject, or alter script's execution \cite{attacks}. Web scripts execute a range of actions, from simple tasks like locating and changing document object model (DOM) elements to harmful tasks such as injecting dynamic code or stealing sensitive user data \cite{martin,toreini2019}. With the increasing complexity of today's web applications and the development of sophisticated evasion techniques, conventional detection methods are not enough \cite{Wang2024}. Signature-based detection and anomaly-based detection with machine learning (ML) classifiers are the conventional approaches which are usually not feasible when large datasets are not available, requiring better interpretable and adaptable classification techniques \cite{sigma}.

In order to address these challenges, this paper presents an automaton-based classification framework for analyzing script behaviors. Script behaviors refer to the observable actions performed by JavaScript code during its execution, which collectively reveal the script's functionality and potential for malicious activity. The classification of script behavior is a fundamental step in improving web security, as it allows a deeper understanding of how attacks work \cite{behavior}. Understanding attack behavior supports the detection of new versions or variances of malware scripts, facilitating the development of generalized defenses \cite{Ma2012}. Analyzing malicious scripts through behavior modeling makes it possible to find common patterns that all variants of each attack type share \cite{JING2022193}. Then, the script's maliciousness can be determined by comparing its execution trace to established behavior models of known attacks.

The developed system uses scripts to model execution behaviors through sequences that demonstrate particular activities, such as DOM element generation, storage operations or dynamic code injection. By analyzing these sequences, it is possible to assess the degree to which the observed behaviors are aligned with a set of known malicious patterns. This is performed with a Deterministic Finite Automaton (DFA) system that tracks state transitions by monitoring observed script behaviors is constructed. This system provides a deterministic way to display action sequences so that it can identify whether scripts are benign, malicious, or partially malicious.

A key innovation of the system is the use of behavior weights, which assign significance to each script action based on its importance in an attack context. This weighting system ensures a finer-grained risk assessment by recognizing that certain individual behaviors inherently carry more risk than others. For example, actions like accessing cookies or interacting with the DOM may typically be benign, while behaviors such as injecting dynamic code or sending data are probably more suspicious due to their potential use in malicious activities. This weighting system has been developed in close collaboration with JavaScript experts, ensuring that the assigned weights accurately reflect the real-world threat dynamics. By assigning higher weights to high-risk behaviors, the system prioritizes their significance in the classification process, enabling a more accurate evaluation of scripts with potential threats.

Therefore, this work proposes a novel classification system that combines automata theory and behavior weighting. Unlike traditional classifiers, the proposed method offers structured and interpretable decision making, increasing transparency in script behavior analysis. Experimental evaluation shows that this method is effective in distinguishing benign from malicious scripts, opening a promising path to improve automated threat detection systems. Through this work, our goal is to complement existing solutions by offering a behavior-based classification system that enhances web application security against evolving threats.

This paper is divided into several sections that provide details that could help researchers replicate the baseline and compare it with their findings. \hyperref[sec:2]{Section 2} presents related works. \hyperref[sec:3]{Section 3} details the preparation steps and the methodology used to implement our system. \hyperref[sec:4]{Section 4} provides an in-depth analysis of the results obtained in a proprietary dataset. Finally, \hyperref[sec:5]{Section 5} reviews the primary conclusions reached and suggests potential directions for future research.

\section{Related Work}\label{sec:2}

The detection of JavaScript malware has relied on static, dynamic, and hybrid analysis techniques, each with inherent limitations. Static analysis approaches such as ZOZZLE \cite{zozzle} and PROPHILER \cite{Prophiler} seek to predict maliciousness from syntactic and lexical features extracted from JavaScript code. For example, ZOZZLE detects obfuscated malware by using statistical models over Abstract Syntax Trees, while PROPHILER uses lightweight static features to classify scripts. However, these methods often fail at highly obfuscated or dynamically generated scripts which do not present themselves as obvious static patterns \cite{zozzle,Prophiler,jsand}. 

Dynamic analysis tools, such as JSAND \cite{jsand}, execute JavaScript in controlled environments to extract behavioral features like re-directions, de-obfuscation, and exploit attempts. This method detects unseen threats through runtime behavioral analyzes, as seen in the work of Gorji et al. \cite{gorji} who clustered predictive function calls to create behavioral signatures. However, this analysis technique is not only computationally expensive, but also vulnerable to evasive malware that can detect and escape sandbox testing environments. Tools like FV8 \cite{FV8} have been built to detect evasive actions in malicious scripts. FV8 changes the V8 JavaScript runtime in order to force code execution on API's that inject dynamic code, increasing the code coverage and visibility of malicious behaviors.

Hybrid analysis methods, such as CUJO \cite{CUJO} employ static and dynamic features, but still face scalability issues and difficulties in modeling sophisticated behavioral sequences. He et al. \cite{he2018} developed a combined static and dynamic analysis system that uses Random Forest classifier to analyze feature-rich datasets achieving high detection accuracy. The context-sensitive characteristics of JavaScript malware remain a challenge for hybrid analysis approaches despite recent improvements.

The integration of ML techniques within JavaScript malware detection has significantly improved the detection process, showing superior results compared to traditional methods \cite{review}. A systematic review of the literature on studies from 2009 to 2019 reveals that ML-based detection models achieve greater precision in detecting malicious JavaScript code \cite{review}, underscoring the growing reliance on ML for malware detection. For example, Jodavi et al. \cite{jodavi} utilized ensemble of support vector machine classifiers with a binary particle swarm optimization algorithm to identify vulnerable JavaScript code, reporting f-measure over 86\%. However, despite these promising results, ML-based approaches have several inherent limitations. Successful operation of these models depends on extensive datasets and faces challenges when addressing new attacking methods \cite{review}. The black-box operational nature of numerous ML algorithms leads to complicated understanding of detection decision \cite{ribeiro2016} by analysts who need to interpret the rationale supporting classification results, thus diminishing practical usability in dynamic security environments.

In contrast, automata-based models offer a structured and transparent alternative by explicitly modeling the sequence of actions performed by scripts during execution. Automata-based models have been recently explored as ways to represent the malware's behavior. Previous studies have used Finite State Automaton (FSA) to represent the behavior of processes based on system calls sequences \cite{605929,layered}. For example, Sekar et al. \cite{Sekar} used heuristic-based methods to describe normal behavior using FSAs, but the approach is limited to heuristics that are predefined. Xue et al. \cite{Xue} introduced a DFA to abstract and summarize common behaviors of malicious JavaScript associated with specific attack types, effectively identifying variants of known malware through behavioral patterns rather than static signatures. This makes them particularly useful for detecting obfuscated or polymorphic malware. The results of these studies indicate a shift towards modeling the dynamic execution patterns of malware in order to improve detection capabilities. Lemberger et al. \cite{RegisterAutomata} explored the use of register automata and pushdown systems to describe malware specifications, in order to capture more complex and context sensitive interactions in execution traces. These models extend beyond DFA's by incorporating program stack and register values, though they still face limitations in addressing the dynamic and adaptive nature of JavaScript malware.

Therefore, to overcome the limitations of current detection methods in handling context-based behaviors and the dynamic nature of JavaScript malware, this work introduces a novel approach focused on enhancing script behavior analysis. Although AI techniques have shown some advantages in certain cases, their practical application in this project is not the most appropriate. In addition to requiring a large amount of data, these methods act as 'black boxes, making their decision-making processes opaque and difficult to interpret. In environments where transparent and timely analysis is key, lack of clarity in AI results can impede rapid response and expert assessment. This work enhances script behavior analysis by investigating sequential and contextual behavior patterns during execution, aiming to identify characteristics that distinguish benign from malicious JavaScript activities, thereby providing enhanced adaptability, robustness to emerging risks, and clarity in how decisions are made.

%Our proposed models aim to detect and classify malicious JavaScript by analyzing execution patterns in detail, focusing on how specific behaviors progress, and assessing their effects on attack vectors to establish behavioral characteristics that distinguish benign from malicious activities.

\section{Methodology}\label{sec:3}

To effectively detect malicious JavaScript sequences, we begin by analyzing the behaviors that scripts exhibit during execution. Our research started by analyzing a proprietary dataset which represents JavaScript execution details using structured attributes including behavior identifiers alongside timestamps and execution contexts and DOM interactions. The records include information about the invoked methods, their associated behaviors (e.g. \textit{"Add DOM Element(s)"}, \textit{"Access Cookies"}) and metadata about the element and script involved, labeled as malicious or benign. From this comprehensive dataset, we extract sequences of behaviors, which are organized into a relational database for further analysis and DFA construction. It should be noted that this research relies on our unique dataset because no existing public dataset provides equivalent dynamic execution details, making our proprietary dataset essential for this study.

Each sequence in the dataset is identified by a unique ID and is represented as an ordered list of behaviors. These behaviors capture key execution events, such as DOM manipulations (\textit{"Find DOM Element(s)"}, \textit{"Update DOM Element"}, \textit{"Inject Code Dynamically"}), storage access (\textit{"Access Browser Storage"}, \textit{"Access Cookies"}, \textit{"Change Cookies"}), and network interactions (\textit{"Send Data"}). Each script behavior receives a weight value that determines how important it is to detect benign and malicious behaviors. The weight system assigns \textit{"Find DOM Element(s)"} and similar behaviors lower values because their occurrence is harmless, typically doesn't exhibiting malicious behavior. On the other hand, the system assigns elevated weights to dangerous behaviors such as \textit{"Inject Code Dynamically"} or\textit{ "Send Data"} because they act as important elements for carrying out malicious activity. This assignment reduces the influence of low-risk behaviors while amplifying high-risk behaviors to create sophisticated risk assessments through action weights in the evaluation process. The label of each sequence, indicating whether it is malicious or benign, is derived directly from the dataset.

To illustrate how the behavior weighting system works, let's consider two examples, one representing a benign script behavior and another demonstrating a malicious behavior. For a benign example, let's consider a script that performs routine DOM manipulations to dynamically update a webpage. The sequence of actions begins with locating a DOM element, such as a button or a div, represented by the behavior \textit{"Find DOM Element(s)"} that has a weight of 2. The script then modifies element attributes through \textit{"Update DOM Element"} which has a weight value of 3. To conclude, the script introduces fresh DOM content through element additions like banners or notifications which receive a weight value of 1. In our dataset this sequence will be represented as [\textit{"Find DOM Element(s)"},\textit{"Update DOM Element"},\textit{"Add DOM Element(s)"}]. This sequence has the associated behavior weights of 2, 3 and 1 respectively. These standard activities have received low weight values because they maintain typical behavior patterns without creating significant risk.  

For a malicious example inspired by a Magecart-style attack, we usually have behaviors associated with credit card skimming and data theft \cite{rus2023defeating}. At the start of this sequence, the script normally adds harmful code to the webpage, represented by the \textit{"Inject Code Dynamically"} behavior with a weight of 4. The injected code performs searches for sensitive form fields through \textit{"Find DOM Element(s)"}, associated with a weight value of 2. Once these elements are identified, the script captures user data through the behavior \textit{"Access Input"}, assigned a weight of 4, and subsequently sends the stolen data to an external server using \textit{"Send Data"} with a weight of 5. The sequence [\textit{"Inject Code Dynamically"}, \textit{"Find DOM Element(s)"}, \textit{"Access Input"}, \textit{"Send Data"}], associated with the weights of 4, 2, 4 and 5 respectively, demonstrates a systematic progression from start to finish of malicious activities. The sequences risk level increases through weight assignments that focus on actions such as code injection and data exfiltration, which distinguish it from routine system behaviors.

To model the behavior of malicious JavaScript scripts, a DFA, called behavior DFA is built. Formally, a DFA is defined as a 5-tuple \( (Q, \Sigma, \delta, q_0, F) \) where \( Q \) is the set of states, \( \Sigma \) is the input alphabet, \( \delta \) is the transition function from \( Q \) × \( \Sigma \) to \( Q \), \( q_0 \) is the initial state and F is the set of final (accepting) states. In the context of malicious JavaScript modeling, the state space \( Q \) represents the distinct states that a script can transition through during execution. These states encode significant intermediate conditions of the execution sequence, such as DOM manipulation, data exfiltration, or script injection. The input alphabet
\( \Sigma \) consists of the possible execution behaviors, such as \textit{"Find DOM Element(s)"}, \textit{"Inject Code Dynamically"}, and \textit{"Send Data"}. Transitions between states, governed by \( \delta \), occur based on the sequence of observed script behaviors.

\begin{figure}[H]
    \centering
    \includegraphics[width=0.9\linewidth]{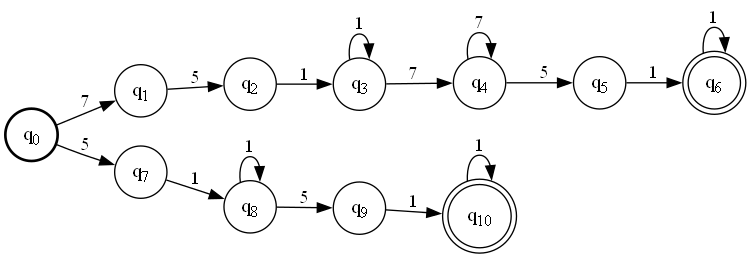}
    \caption{Behavior DFA example.}
    \label{fig:enter-label}
\end{figure}

The construction of the behavior DFA starts with an analysis of existing malicious JavaScript behavioral sequences. Identified sequences serve as basic elements in creating state transition definitions that help to recognize states with malicious behavior characteristics. Each execution behavior corresponds to a transition between states, where the path through the behavior DFA captures the sequence of actions. Any sequence that reaches a final state F, where malicious execution occurs, is flagged as a potential threat. \hyperref[fig:enter-label]{Fig. 1} present a visual representation of the proposed behavior DFA system constructed using two JavaScript behavior sequences: 

[\textit{"Add Event Handler"}, \textit{"Set Callback"}, \textit{"Find DOM Element(s)"}, \textit{"Find DOM Element(s)"}, \textit{"Find DOM Element(s)"}, \textit{"Add Event Handler"}, \textit{"Add Event Handler"}, \textit{"Set Callback"}, \textit{"Find DOM Element(s)"}, \textit{"Find DOM Element(s)"}, \textit{"Find DOM Element(s)"}] and

[\textit{"Set Callback"}, \textit{"Find DOM Element(s)"}, \textit{"Find DOM Element(s)"}, \textit{"Find DOM Element(s)"}, \textit{"Set Callback"}, \textit{"Find DOM Element(s)"}, \textit{"Find DOM Element(s)"}, \textit{"Find DOM Element(s)"}, \textit{"Find DOM Element(s)"}]. 

These encoded sequences will result in [7, 5, 1, 1, 1, 1, 7, 7, 5, 1, 1, 1] and [5, 1, 1, 1, 5, 1, 1, 1, 1], respectively. Each transition in the behavior DFA is labeled with a behavior represented by an identifier, 7 for example corresponds to \textit{"Add Event Handler"}. The state space \( Q \) consists of labeled states \( q_0 \), \( q_1 \), ..., \( q_{10} \), representing progression through the execution sequence. The behavior DFA starts in the initial state \( q_0 \), and transitions between states occur based on the detected behavior having two final states \( q_6 \) and \( q_{10} \).

To classify a new set of behaviors, the behavior DFA evaluates the sequence of actions performed during the JavaScript execution. The behavior DFA receives sequences of states as transitions that begin from its initial state \( q_0 \). During sequence processing, the behavior DFA advances through states according to input behaviors which follow the behavior DFA defined transitions. The behavior DFA detects potential maliciousness whenever a sequence reaches a final state and results in a partially malign behavior when it ends in a non-final state.

As it is essential to identify not only exact matches to known malicious patterns but also taking into account sequences that may share similar behaviors, we further refine the classification process using a distance to the nearest final state. We compute the distance to the closest final state using breadth-first search (BFS) when a sequence does not perfectly match any known malicious pattern. This approach calculates the shortest path from the sequence's last state to a confirmed malicious state, with each transition accumulating a cost based on behavior weights. Lower distances indicate that a sequence is more similar to a known malicious pattern, while higher distances suggest that the sequence is less likely to be malicious. For instance let's consider a sequence [7, 5]. Using the behavior DFA from \hyperref[fig:enter-label]{Fig. 1} we can see that this sequence ends in a state that does not directly reach a confirmed malicious state. Using BFS, we calculate the cost of the shortest path from the sequence's last state, in this case \( q_2 \), to all malicious state. In this case, the closest final state is \( q_6 \).

Then the classification algorithm will give us the probability of a given sequence being malicious. To do that, we calculate the match percentage, which is the ratio of how well the sequence matches known malicious patterns. The formula for match percentage is: \[ \text{Match Percentage} = \left( \frac{\text{Total Matched Behavior Weight}}{\text{Total Weight to Nearest Final State}} \right) \times 100 \] 

In this formula, the "Total Matched Behavior Weight" represents the sum of transition weights that lead to the current state. The "Total Weight to Nearest Final State" represents the sum of transition weights which directly lead to the nearest known malicious state, starting on the initial state. For example let´s consider the sequence [7, 5], knowing that behaviors 7 and 5 have a weight of 3. Comparing the sequence with the behavior DFA in \hyperref[fig:enter-label]{Fig. 1} we can see that this sequence matches the behavior DFA, ending in state \( q_2 \). The sum of behaviors weight matched, 7 and 5, will be the "Total Matched Behavior Weight", in this case 3 + 3, that is, 6. From this state the nearest final state is \( q_6 \), thus the "Total Weight to Nearest Final State" in this case will be the sum of all behaviors weight from state \( q_0 \) to \( q_6 \) [7, 5, 1, 7, 5, 1]. Assuming that behavior 1 has a weight of 2, this results in the following sum, 3 + 3 + 2 + 3 + 3 + 2 = 16. Applying the formula, we have 6 / 16 * 100, which will give us the results of 37.5\%. 

We classify a sequence as malicious if it matches a known attack pattern perfectly, partially malicious if the sequence matches some behaviors associated with malicious activity, or benign if there is no strong similarity. For partially malicious sequences, the classification includes a match percentage, reflecting the degree of similarity to known malicious patterns. In this case, the sequence is classified as partially malicious with a match percentage of 37.5\%. This result indicates that the script exhibits some suspicious behaviors but has not yet fully demonstrated malicious intent. It is necessary to perform additional behavioral analysis and monitoring to assess any potential threat from the script.

The proposed approach employs known attack patterns to construct itself, but its ability to classify behaviors partially enables it to detect unknown malicious actions that are similar to existing attacks, since diverse attacks use similar behavioral sequences. This is consistent with behavior-based detection frameworks as demonstrated by Ravi et al. \cite{Ravi2013}, where partial behavioral overlaps, common in adversarial campaigns due to shared tactics such as privilege escalation and code injection allowed one to generalize novel threats. The behavior DFA benefits from this shared behavior pattern between known attacks to detect malicious actions that are similar from recognized patterns, thereby strengthening its detection efficiency. Moreover, the proposed approach provides straightforward mechanisms for adding newly discovered malicious sequences. Through its adaptable design, the model maintains operational effectiveness by easily updating its threat detection capabilities without necessitating a complete redesign.

\section{Results and Evaluation}\label{sec:4}

The evaluation criteria for this work focus on classifying sequences based on their match percentage, which reflects how closely a sequence aligns with the defined malicious patterns within the behavior DFA, using the formula shown in \hyperref[sec:3]{Section 3}. A sequence is classified as malign if it exhibits a perfect match with a known malicious pattern, as partially malign if it shows partial alignment, or as benign if no significant alignment is observed. The behavior DFA was initially built using known malicious sequences and subsequently tested on a proprietary dataset. The evaluation methodology combines real-world data assessment with a solid system to detect benign from malicious activities.

The system evaluated 1058 sequences, classifying 10 as malign, 288 as partially malign and 760 as benign. The results are consistent with expectations, since the behavior DFA was built from 10 malicious sequences. This design guarantees that the behavior DFA can detect known malicious patterns with high precision and can discriminate sequences that have partial similarities. Further categorization of the partially malign sequences found match percentages of 18.75\% (165 sequences), 30\% (73 sequences), 37.5\% (25 sequences) and 50\% (25 sequences) in the proprietary dataset. These variations show how closely aligned sequences are to different malicious patterns, demonstrating that the system can capture partial matches, demonstrating the importance of analyzing sequences that share sufficient similarities with existing patterns. Attack sequences frequently display shared behaviors including like event handler addition and DOM element interaction, but its important to note that these behaviors can also exist within benign sequences.

To demonstrate the functionality of the classification approach, we present a case study for the sequence ID 1058 containing the next sequence of behaviors: [7, {11, 3}, 7, {11, 3}]. Each number in the sequence represents an identifier for a behavior, where, for example, 7 represents \textit{"Add Event Handler"}, 11 represents \textit{"Send Data"}, and 3 represents \textit{"Update DOM Element"}. This sequence indicates that a script has implemented actions to maintain continuous communication and data extraction methods. The initial stage involves the addition of event handlers that could serve either for user monitoring or interaction response purposes. The script follows these actions with data transmission and DOM element modifications, which indicates potential information exfiltration or preparation for extra malicious activity like dynamic injections or deceptive content display. Each of these script actions keeps repeating, which strengthens the hypothesis that the script attempts to sustain ongoing communication and repeated user interface modifications for interaction manipulation. However, this sequence is classified as benign on our dataset, the repetition of adding event handlers and subsequent interactions with the DOM suggests that the script can also be used to perform legitimate actions, such as dynamically updating page content or enabling user interactions. The inclusion of \textit{"Send Data"} may correspond to benign functionalities, such as analytics tracking or communicating with a server for non-malicious purposes. So, it can be interesting to test this sequence on our behavior DFA.

When processed by the behavior DFA, the match percentage was calculated as 18.75\%, classifying the sequence as partially malign. The sequence 1058, matched the behavior DFA only on the first behavior (7). The nearest final state found was the state \( q_{25} \), with the sequence [7, 5, 1, 7, 5, 1,] representing the path from the state \( q_{0} \) to this state, \( q_{25} \). Knowing that behavior 7 and behaviors 5 had weights of 3 and behavior 1 has a weight of 2 we can calculate the final match percentage. The total matched behavior weight is 3 representing the matched behavior 7 and the total weight to the nearest final state is the sum of the weights for the sequence leading to state \( q_{25} \), \(3 \,  + 3 \,  + 2 \,  + 3 \,  + 3 \,  + 2 \,  = 16.\) The final match percentage is then calculated as:
\[
\frac{\text{Total Matched Behavior Weight}}{\text{Total Weight to Nearest Final State}} \times 100 = \frac{3}{16} \times 100 = 18.75\%.
\]
\medskip

The partial match outcomes demonstrate the behavior DFA's capability to analyze sequences, offering nuanced insights into their alignment with known malicious patterns. The behavior DFA produces partial matches that measure the varying degrees of similarity to known attack sequences. The partially malign sequences were further analyzed based on their match percentages, revealing different degrees of alignment with malicious behaviors and the importance of analyzing them thoroughly:
\begin{itemize}
    \item 18.75\% (165 sequences): 165 sequences of the dataset were classified as partially malign with a match percentage of 18.75\%. They matched the behavior DFA on behavior 7 (\textit{"Add Event Handler"}, weight: 3) with a total weight to the nearest final state of 16. The presence of this behavior points to script a functionality that mostly serve as standard user interface interactive mechanisms. However in malign scenarios its appearance may also represent a preparatory step in a broader malicious strategy if followed by more critical actions. The sequences show relatively low match percentage with established malicious patterns, but suggest a monitoring need for this emerging suspicious activity.

    \item 30\% (73 sequences): In the entire dataset 73 sequences were classified as partially malign with a match percentage of 30\%. These sequences matched the behavior DFA on behavior 5 (\textit{"Set Callback"}, weight: 3) with a total weight to the nearest final state of 10. This indicates a script executing activities which can range from harmless preparation steps to benign tasks that connect event with callbacks for user interaction management or functionality activation on web pages. However, setting a callback can also be exploited in malicious contexts, where it serves as a step in executing further harmful actions, such as tracking user input or hijacking event-driven functionality. The absence of others high-risk behaviors in this sequence leans toward a benign classification, though the potential for future malicious activity justify a detailed analysis.
    
    \item 37.5\% (25 sequences): A total of 25 sequences in the dataset were identified as partially malign, exhibiting a 37.5\% similarity threshold to known malicious patterns. In these sequences the behaviors matched were 7 and 5 (\textit{"Add Event Handler"} and \textit{"Set Callback"}, weights: 3 each) with a total weight to the nearest final state of 16. The addition of an event handler suggests the script is interacting more closely with the user or the DOM. This could be entirely benign, such as enabling custom events to improve user experience or monitor page functionality. On the other hand, in a malign scenario, adding event handlers may be used to monitor sensitive user inputs or trigger harmful actions dynamically in response to user interactions. The sequence's match percentage reflects its potential to be either benign or malicious, depending on its broader execution context.
    
    \item 50\% (25 sequences): Of the total dataset, 25 sequences were classified as partially malign with a match percentage of 50\%. From these sequences 19 matched behaviors 5 and 1 (\textit{"Set Callback"} and \textit{"Find DOM Element"}, weights: 3 and 2, respectively) with a total weight to the nearest final state of 10. These behaviors are often benign, as scripts frequently search for DOM elements and set callbacks for event-driven programming. For example, they might be part of a legitimate script dynamically loading content or enhancing interactivity. However, the relatively high match percentage with malicious patterns signals caution, as these same actions could also be used to locate elements for data exfiltration or as part of a setup to inject harmful code. The classification as partially malign highlights its dual potential. Other 6 sequences matched behaviors 7, 5 and 1 (\textit{"Add Event Handler"}, \textit{"Set Callback"} and \textit{"Find DOM Element"}, weights: 3, 3 and 2, respectively) with a total weight to the nearest final state of 16. For benign activity, these actions might represent routine DOM manipulation and event management to enable dynamic functionality. For instance, adding event handlers and setting callbacks could be part of a framework managing user inputs. The same sequence of actions can lead to severe attacks in malign situations because it indicates vulnerabilities exploitation and targeted payload delivery against specific DOM elements. The 50\% match rate indicates that script behavior analysis is essential to distinguish between attacks with benign intent and those with malicious objectives.
\end{itemize}

Overall, the system successfully shows its ability to detect different malicious pattern alignments, measuring effectively diverse threat levels from partially matching behaviors. Using the partial match functionality as its key strength, the system is able to identify exact and similar patterns from known attacks, remaining effective against evolving attack approaches. Application match percentage calculations from the system provide essential security information regarding partly suspicious sequences, supporting early threat detection without unnecessary false alarms. The findings align with the behavior DFA's design, based on 10 malicious sequences, ensuring the identification of attack components while retaining sensitivity to partial overlaps. This enables operators to flag suspicious activities even when attacks exhibit only partial similarity to known threats to known threats, as reviewed and validated by domain experts.

\section{Conclusion}\label{sec:5}
This study presents a behavior-based approach to detect malicious JavaScript through a behavior DFA combined with behavior weighting. By modeling script execution as sequences of weighted actions, the system detects both exact and partial matches based on known malicious patterns, identifying evolving threats. The weighting system ensures that high-risk behaviors, such as code injection or data exfiltration, have greater impact on the classification process. Even though typical benign actions are also taken into account, they have less impact on the final classification.

The evaluation process demonstrated the system's ability to generalize beyond its known malicious pattern, classifying sequences as benign, partially malicious, or fully malicious based on a match percentage that represents their similarity to known malicious states. This capability highlights the adaptability of the system, as it can recognize new threats that share behavioral similarities with known attacks. Compared to traditional AI techniques, this approach offers greater flexibility, resilience against new threats, and transparency in the decision-making process. The modular design structure of the behavior DFA enables the improvement of its set of known attacks with new attack patterns, preserving its ongoing utility against new malware approaches, without requiring complete retraining.

The next steps to improve over this work are focused on the implementation of subpattern detection in the behavior DFAs to enhance it's capabilities. By identifying malicious subquences embedded within larger, seemingly benign workflows, the system's granularity and sensitivity will be improved. This will involve redesigning the automaton along with matching logic reset functionality at each input step, letting the system identify modular attacks wherever they occur in the sequence. Future development will also include the implementation of external elements which use ML models or API's to examine scripts' visited URLs and resources. The integrated system will enhance threat detection capabilities through its ability to identify malicious redirects or compromised external resources. These improvements will further contribute to the security and robustness of web applications against the evolving JavaScript-based threats.

\textbf{Acknowledgements.}
This work was done and funded in the scope of the BEHAVIOR project (NORTE2030-FEDER-00576300 no. 14391). This work was also supported by UIDB/00760/2020.

\bibliographystyle{splncs04}
\bibliography{refs.bib}
\end{document}